\documentclass{nature}[a4paper]
\usepackage{bm}
\usepackage{pifont}
\usepackage{marvosym}
\usepackage[colorlinks,linkcolor = blue,citecolor = blue,anchorcolor = blue]{hyperref}
\usepackage{multirow}
\usepackage{geometry}
\usepackage{authblk}
\usepackage{amsmath}
\usepackage{graphicx}
\usepackage{xcolor}
\makeatletter
\let\saved@includegraphics\includegraphics
\AtBeginDocument{\let\includegraphics\saved@includegraphics}
\renewenvironment*{figure}{\@float{figure}}{\end@float}
\makeatother
\usepackage[labelfont=bf]{caption}
\usepackage{cite}
\usepackage{soul}

\title{High spin axion insulator \footnotetext{$^\ast$  E-mail: liyuhang@nankai.edu.cn, jianghuaphy@fudan.edu.cn}}

\author[1,2]{Shuai Li}
\author[3]{Ming Gong}
\author[4\ \Letter]{Yu-Hang Li}
\author[5,2,7,8\ \Letter]{Hua Jiang}
\author[3,5,6,7,8]{X. C. Xie}

\affil[1]{School of Physical Science and Technology, Soochow University, Suzhou 215006, China}
\affil[2]{Institute for Advanced Study, Soochow University, Suzhou 215006, China}
\affil[3]{International Center for Quantum Materials, School of Physics, Peking University, Beijing 100871, China}
\affil[4]{School of Physics, Nankai University, Tianjin 300071, China}
\affil[5]{Institute for Nanoelectronic Devices and Quantum Computing, Fudan University, Shanghai 200433, China}
\affil[6]{Hefei National Laboratory, Hefei 230088, China}
\affil[7]{Interdisciplinary Center for Theoretical Physics and Information Sciences (ICTPIS), Fudan University, Shanghai 200433, China}
\affil[8]{Institute for Nanoelectronic Devices and Quantum Computing, Fudan University, Shanghai 200433, China}

\graphicspath{{figures/}}

\date{}                   
\begin{document}
	
	\maketitle
	
	\vspace{10pt}
	
	\begin{abstract}
		\textbf{Axion insulators possess a quantized axion field $\theta=\pi$ protected by combined lattice and time-reversal symmetry, holding great potential for device applications in layertronics and quantum computing. Here, we propose a high-spin axion insulator (HSAI) defined in large spin-$s$ representation, which maintains the same inherent symmetry but possesses a notable axion field $\theta=(s+1/2)^2\pi$. Such distinct axion field is confirmed independently by the direct calculation of the axion term using hybrid Wannier functions, layer-resolved Chern numbers, as well as the topological magneto-electric effect. We show that the guaranteed gapless quasi-particle excitation is absent at the boundary of the HSAI despite its integer surface Chern number, hinting an unusual quantum anomaly violating the conventional bulk-boundary correspondence. Furthermore, we ascertain that the axion field $\theta$ can be precisely tuned through an external magnetic field, enabling the manipulation of bonded transport properties. The HSAI proposed here can be experimentally verified in ultra-cold atoms by the quantized non-reciprocal conductance or topological magnetoelectric response. Our work enriches the understanding of axion insulators in condensed matter physics, paving the way for future device applications.}\\
	\end{abstract}

\noindent Symmetry plays an essential role in understanding the behavior of condensed materials~\cite{wooten_2008,Goldhaber2003Symmetry,zee2015fearful,McGreevy2023Generalized}. For example, in the presence of time-reversal symmetry, three dimensional insulator typically falls into two categories: one is trivial insulator while the other is topological insulator~\cite{Qi2011Topological,Hasan2010Colloquium}. These divergent categories can be well described within the framework of the Chern-Simons theory, where the Lagrangian incorporates an additional symmetry allowed term $\mathcal{L}_{\theta}=\int{dtd{\bf{r}}^3}\alpha\theta/(4\pi^2){\bf{E}}\cdot{\bf{B}}$ with ${\bf{E}}$ and ${\bf{B}}$ the conventional electric and magnetic fields, $\alpha$ the fine structure constant, and $\theta$ the gauge dependent axion term~\cite{Qi2008Topological}. Because of the $2\pi$ periodicity under a gauge transformation~\cite{Vazifeh2010Quantization}, the axion term here is well defined within the region $[0,\ 2\pi)$. Besides, as the quantity ${\bf{E}}\cdot{\bf{B}}$ flips sign under time-reversal ($\mathcal{T}$) operation, the axion field manifests only two distinct values, that is, $\theta=0$ for normal insulator and $\theta=\pi$ for topological insulator~\cite{Qi2008Topological}. Furthermore, the non-vanishing axion term in the Lagrangian would introduce additional magneto-electric responses to the Maxwell equations and in turn, results in a distinctive topological magneto-electric effect~\cite{armitage2019matter,Litvinov2020,planelles2021axion}, furnishing a hallmark to the quantized axion field.

\noindent In addition to the time-reversal ($\mathcal{T}$) symmetry, the quantized axion field $\theta=\pi$ can also be protected by combined lattice and time-reversal symmetry (for example $\mathcal{S}=\mathcal{T}\tau_{1/2}$ with $\tau_{1/2}$ the half translation operator)\cite{Varnava2018Surface}, as the quantity ${\bf{E}}\cdot{\bf{B}}$ undergoes the same sign reversal. This kind of materials, termed axion insulator~\cite{Wang2022Topological,Sekine2021Axion,Zhao2021Routes,Tokura2019Magnetic,Nenno2020Axion,Wang2015Quantized,Morimoto2015Topological,Mogi2017A,Xiao2018Realization,Xu2019Higher}, holds significant potential in layertronics~\cite{Gao2021Layer,Gong2023Half,Chen2022Layer,Dai2022Quantum,Li2023Dissipationless} and quantum computing~\cite{Monica2019Visualization,Yan2021A}. MnBi${}_2$Te${}_4$ and its family have recently been proposed as axion insulators in the antiferromagnetic state~\cite{Wang2022Topological,Li2022Identifying,Zhang2019Topological,Li2019Intrinsic,Liu2020Robust,Otrokov2019Prediction}, which finds a concise description with an effective Hamiltonian defined on the orbital and spin-$1/2$ spaces~\cite{Zhang2019Topological}. Because the symmetry transformations of high spin representations and spin-$1/2$ are identical (see Supplementary Section 1), in this work, we generalize this model to other spin species and thus propose a high spin axion insulator (HSAI) preserving the same symmetry. We find that the HSAI with spin-$s$ possesses a notable axion field $\theta_{HSAI}=(s+1/2)^2\pi$. It carries a multiple quantized helical hinge current (QHHC) that is robust against disorders even in the absence of the topologically protected gapless exciations, which contradicts the integer surface Chern number. Consequently, HSAI exhibits an unusual quantum anomaly that violates the conventional bulk-boundary correspondence. In contrast to the case of spin-$1/2$ axion insulator, the direct calculation of the axion term shows that the large axion field in high-spin case originates mostly from localized surface Wannier functions while, in the bulk, the axion field is either $0$ or $\pi$. 
Strikingly, we show that the axion field in HSAI can be tuned precisely by manipulating the magnetic configuration through an external magnetic field, providing a pioneering tuning knob to control the QHHC and the quantized magneto-electric response. Possible experimental realization in ultra-cold atoms is also discussed.

\begin{figure}[t]
  	\centering
    	\includegraphics[width=0.8\textwidth]{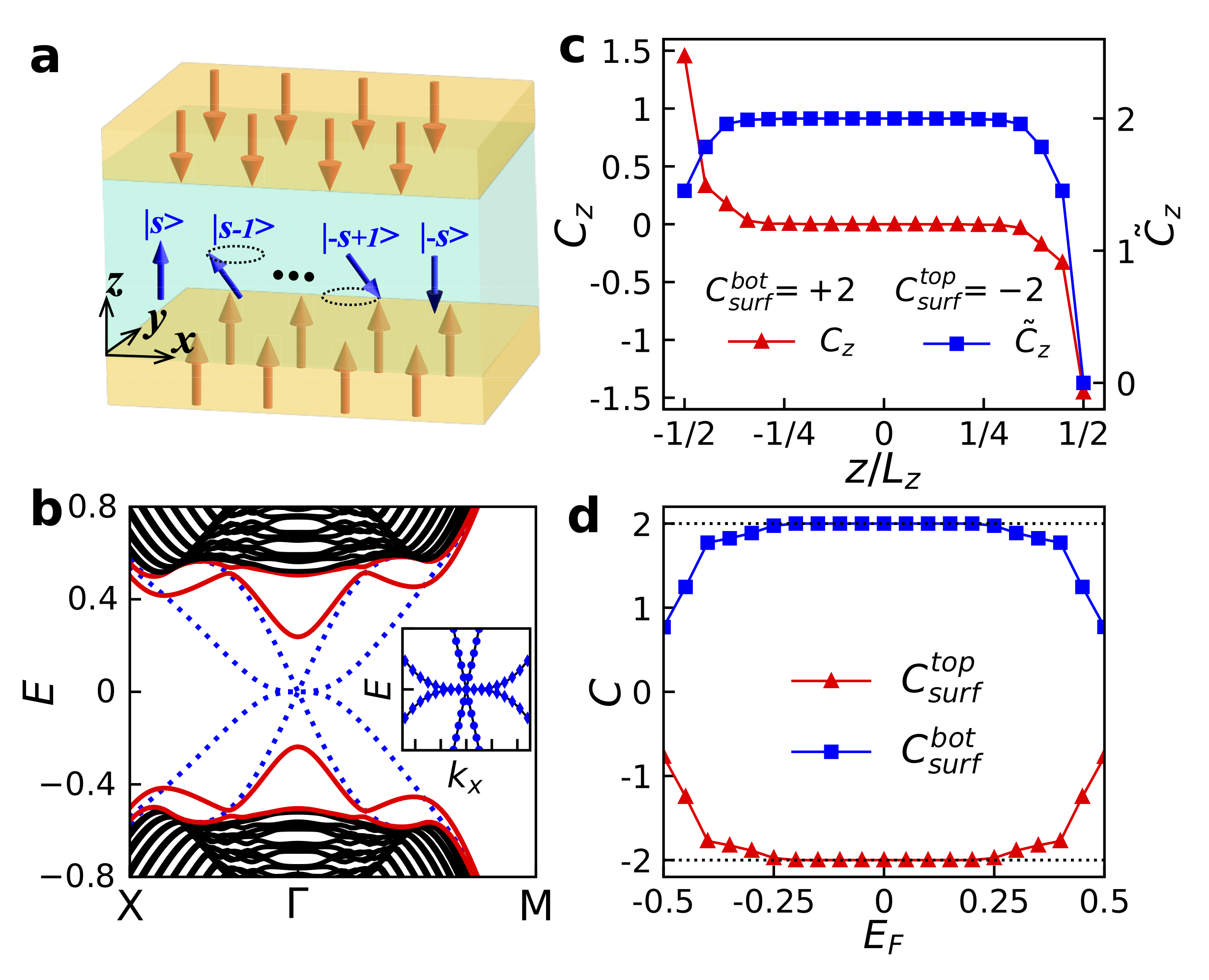}
    	\caption{\textbf{Model of the HSAI.}
		{\textbf{a}} Schematic for the HSAI defined on the $|s,m_z\rangle$ space. The blue arrows represent the electron spin with different magnetic quantum number $m_z$, which takes values ranging from $-s$ to $s$ individually.
		{\textbf{b}} Energy spectra of the spin-$3/2$ HSAI along M$\rightarrow\Gamma\rightarrow$R path on a slab of thickness $L_z$ with  (red solid lines) and without (blue dashed lines) the magnetic exchange interaction. Here, the black lines refer to bulk bands. Inset: Energy dispersion for the spin-$3/2$ HSAI in the absence of magnetic exchange term near the charge neutral point (solid lines) and the fitting data (markers).
		{\textbf{c}} Layer-resolved Chern number $C_z$ and the cumulative Chern number $\tilde{C}_z=\sum_{-L_z/2}^zC_z$ versus the layer index $z$ for the spin-$3/2$ HSAI. The surface Chern numbers $C^{top(bot)}_{surf}$ that summarize the layer-resolved Chern number on the upper (lower) half side is $-2$ ($+2$).
		{\textbf{d}} Surface Chern number as a function of the Fermi energy $E_F$ for the spin-$3/2$ HSAI. 
		Here, the thickness of the HSAI slab is $L_z=20$.
		} 
    	\label{fig1}
\end{figure}

\noindent\textbf{Results}\\
\noindent\textbf{Effective model for the HSAI}\\
\noindent Recalling the effective four-band Hamiltonian for the spin-$1/2$ axion insulator~\cite{Zhang2019Topological}, we consider a generic model defined on the high spin space which can be expressed as
\begin{align}\label{model_Ham}
\mathcal{H}=\sum_{i=0}^3d_i\Gamma_i+\Delta{\bf{m}}_s\cdot{\bf{s}}\otimes\tau_0.
\end{align}
Here, $d_{0,1,2,3}=\begin{bmatrix}m_0-Bk^2,&Ak_x,&Ak_y,&Ak_z\end{bmatrix}$ with $A$, $B$, $m_0$ the system parameters. $\Gamma_{0}=s_0\otimes\tau_3$ and $\Gamma_{i=1,2,3}=s_i\otimes\tau_1$ where $s_i$ and $\tau_i$ are matrices defined on the high spin space and $2\times 2$ orbital space, respectively. The momentum ${\bf{k}}=\begin{pmatrix}k_x,&k_y,&k_z\end{pmatrix}$ is defined on a cubic lattice with the lattice constant $a_0$ inside the first Brillouin zone. This model Hamiltonian is given directly from the spin-1/2 axion insulator. A construction from symmetry perspective is provided in Supplementary  Section 2. It is evident that the first term in Eq.~(\ref{model_Ham}) describes a high-spin topological insulator, which preserves both time-reversal ($\mathcal{T}$) and parity ($\mathcal{P}$) symmetries. The second term describes the exchange interaction between topological electrons and normalized magnetic spins ${\bf{m}}_s=\begin{pmatrix}m_s^x,&m_s^y,&m_s^z\end{pmatrix}$ with $\Delta$ the exchange strength, resembling that in MnBi${}_2$Te${}_4$, hence explicitly breaks the time-reversal symmetry while preserves the $\mathcal{S}$ symmetry in the infinite size limit along $z$-direction. We consider the antiferromagnetic phase of an even-layer slab involving only the antiparallel (or canted) spins in the top and bottom layers as illustrated in Fig.~\ref{fig1}a, which restores the combined parity and time-reversal ($\mathcal{PT}$) symmetry. 
Unless otherwise specified, we adopt the typical model parameters as follows: $A=m_0=1$, $B=\Delta=0.6$, $a_0=1$.

\begin{table}\centering
\begin{tabular}{ |p{1.5cm}||p{1.5cm}|p{1.5cm}|p{1.5cm}|p{1.5cm}|p{1.5cm}|p{1.5cm}|p{1.5cm}|  }
 \hline
 spin-$s$& $\theta_{CS}^{slab}$ & $\theta_{CS}^{surf}$& $\theta_{CS}^{bulk}$& $C_{HWF}^{top}$& $C_{HWF}^{bot}$& $C^{top}_{surf}$& $C^{bot}_{surf}$ \\
 \hline
1/2&$\pi$    &0         &$\pi$& 0     &0      & -1/2 &1/2\\
3/2&$4\pi$  &$4\pi$ &0      & -2    &2      & -2    &2  \\
5/2&$9\pi$  &$8\pi$ &$\pi$& -4    &4      & -9/2 &9/2\\
7/2&$16\pi$&$16\pi$&0     & -8    &8      & -8    &8  \\
$\vdots$& & & & & & & \\
 \hline
\end{tabular} 
\caption{Axion terms and surface Chern numbers for axion insulators with different spins. }
\label{table1}
\end{table}

\noindent Figure~\ref{fig1}(b) displays the two dimensional energy spectra of the spin-$3/2$ HSAI in the absence (blue dashed lines) and presence (red solid lines) of the magnetic exchange term. In the former case, the time-reversal symmetry is present, where the energy spectrum is gapped in the bulk but has two conducting surface states on each side (blue dashed lines). These two surface states can be accurately fitted by a massless Dirac band $E_1\sim k$ and a cubic band  $E_2\sim k^3$ (inset in Fig.~\ref{fig1}b). Turning on the exchange term in the latter case opens a band gap as indicated by the red solid lines in Fig.~\ref{fig1}b. 
In both cases, the energy spectra are doubly degenerated because of the inherent ($\mathcal{T}$ or $\mathcal{PT}$) symmetry.

\noindent\textbf{Layer-resolved Chern number, quantized helical hinge current and quantum anomaly}\\
\noindent To explore the topological properties of the HSAI, we calculate the layer-resolved Chern number $C_z$ along $\hat{z}$-direction~\cite{Li2022Identifying,Wang2021Helical} along with the cumulative Chern number $\tilde{C}_z=\sum_{-L_z/2}^zC_z$. Given the Chern number $C=(s+1/2)^2$ in the odd-layer case~\cite{Lin2022Direct}, the system is a high Chern number insulator as shown in Supplementary Section 6. In the even layer system, the opposite layer-resolved Chern numbers are overall confined antisymmetrically inside few surface  (top and bottom) layers as shown in Fig.~\ref{fig1}c, resulting in a vanishing total Chern number $C=0$. Nevertheless, the surface Chern number on one side turns out to be well quantized [$C^{top(bot)}_{surf}=\mp 2$] when $s=3/2$ as long as the Fermi level lies inside the energy gap (Fig.~\ref{fig1}d).
Because the layer-resolved Chern number is related to the axion field through the relation $\theta_{HSAI}=(C^{bot}_{surf}-C^{top}_{surf})\pi$~\cite{Essin2009Magnetoelectric}. The oppositely quantized surface Chern numbers in spin-$3/2$ HSAI thereby assure a quadruple axion field $\theta_{HSAI}=4\pi$. 
%%%revise later
Moreover, since the Chern number difference between neighboring top (bottom) and side surfaces is an integer, HSAI supports a possible hinge state that is absent in spin-$1/2$ axion insulator~\cite{Gong2023Half}, which allows a subsequent QHHC owing to opposite chiralities on different surfaces. Nonetheless, we find that this QHHC survives counterintuitively without the existence of any hinge state.
%%%

\begin{figure}[t]
  	\centering
    	\includegraphics[width=0.95\textwidth]{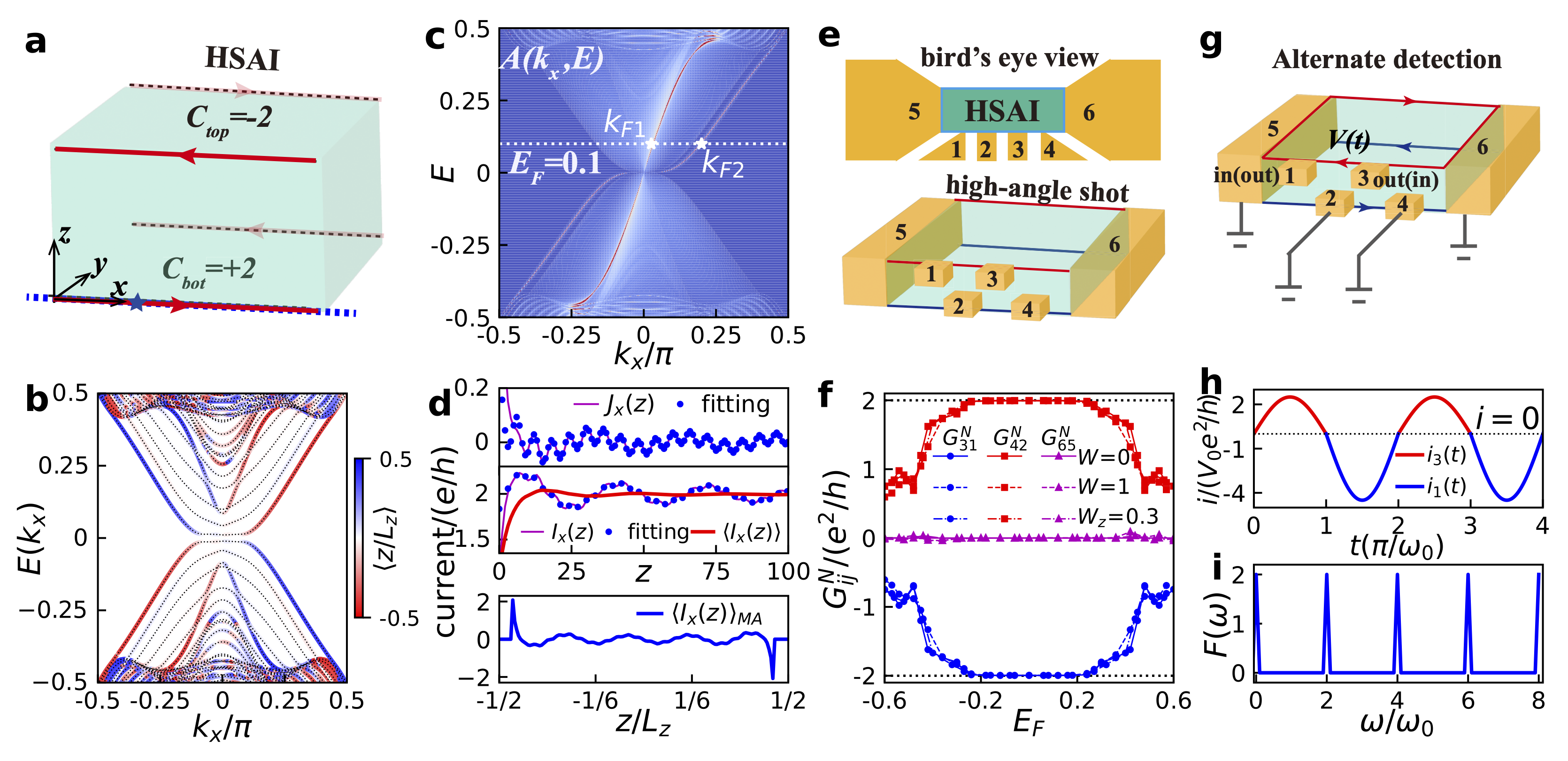}
    	\caption{\textbf{Transport properties of the spin-$3/2$ HSAI.}
	\textbf{a} Schematic current flow in a HSAI. The red arrows denote the quantized helical hinge currents.
	\textbf{b} Energy spectrum and the average position $\langle z/L_z\rangle$ on the front surface for a HSAI nanowire with $L_y=30$, $L_z=16$. 
	\textbf{c} Spectrum density $A(k_x,E)$ for the front lower hinge as labeled by the blue line in \textbf{a} on the $k_x-E$ plane. Here, the system size is $L_y=30$, $L_z=\infty/2$. The white dashed line represents the Fermi energy $E_F=0.1$. The white stars that mark the intersects between the Fermi energy and the spectrum are the Fermi momenta $k_{F1}$ and $k_{F2}$.
	\textbf{d} Top and middle panels are the current density $J_x(z)$, current flux $I_x(z)$ and its $z$-averaged flux $\langle I_x(z)\rangle$ versus the layer index $z$ for a semi-infinite system with size $L_y=30$, $L_z=\infty/2$. The blue dots are the fitting data. Bottom panel shows the distribution of the moving averaged current $\langle I_x(z)\rangle_{MA}$ on the front surface with system size $L_y=30$, $L_z=150$. 
	\textbf{e} Bird's eye view (top panel) and high-angle shot (bottom panel) for the six terminal device.
	\textbf{f} Ensemble-averaged non-reciprocal conductances versus the Fermi energy in the clean limit (W=0), with non-magnetic Anderson disorders of strength $W=1$ and with magnetic Anderson disorders of strength $W_z=0.3$. Here, the system size is $L_x=31$, $L_y=20$, $L_z=21$, and the size of transverse terminals is $10\times10$.
	\textbf{g} Experimental setup to detect the non-reciprocal conductance. In this setup, terminals 2, 4, 5, and 6 are grounded. The voltage is applied alternatively to terminal 1 or terminal 3. 
	\textbf{h} Corresponding temporal dependent current output with parameters $G_{13}=4.5e^2/h$, $G_{31}=2.5e^2/h$.
	\textbf{i} $F(\omega)$ as a function of the frequency $\omega$. 	
	} 
    	\label{fig2}
    	%\vspace{80pt}
\end{figure}

\noindent To clarify this, let us first examine the average position $\langle z/L_z\rangle$ on the front surface of the slab at $y/L_y=-1/2$. The results shown in Fig.~\ref{fig2}b reveals two branches (red lines in Fig.~\ref{fig2}b) concentrated unexpectedly around the bottom hinge ($z/L_z=-1/2$) and propagating rightwards because of the positive group velocity. By contrast, we observe two other branches concentrate oppositely around the top hinge ($z/L_z=1/2$), which, simultaneously, propagate leftwards as depicted by the blue lines. Note that only the results for the front surface (at $y/L_y=-1/2$) are presented here. In the presence of $\mathcal{PT}$ symmetry, the energy spectrum in Fig.~\ref{fig2}b is doubly degenerated as stated above. There are four additional branches existing on the other surface at $y/L_y=1/2$. Because the wavefunctions on the diagonal hinges are connected by this $\mathcal{PT}$ symmetry, they must propagate along the same direction, supporting a helical hinge current. 

\noindent We then turn to the spectrum density $A(k_x,E)$ on a semi-infinite slab~\cite{Gu2021Spectral,Yue2019Symmetry,Chu2011Surface}, where the system extends infinitely along $+\hat{z}$-direction but remains unchanged along the lateral directions. $A(k_x,E)$ on the front lower hinge illustrated by the blue dashed line in Fig.~\ref{fig2}a are plotted in Fig.~\ref{fig2}c. It shows that $A(k_x,E)$ originates mostly from the right-moving energy bands, agreeing remarkably well with the average position in Fig.~\ref{fig2}b. This spectrum density can be verified experimentally by using the nano angle-resolved photoemission spectroscopy and microscopy~\cite{Cattelan2018Perspective,Iwasawa2020High,Brown2014Polycrystalline}. Besides, the high spectrum density on the hinge indicates the presence of a hinge current, the current density of which can be quantitatively determined by~\cite{Chu2011Surface,Gong2023Half}
\begin{align}\label{helical_current}
j_x(E_F,{\bf{r}})=-\frac{e}{h\pi}\int_{-\pi}^{\pi}{dk_x}\text{Im}\{\text{Tr}[\frac{\partial H_{HSAI}(k_x,{\bf{r}})}{\partial k_x}G^r(E_F;k_x,{\bf{r}})]\},
\end{align}
where $E_F$ is the Fermi energy labeled by the white line in Fig.~\ref{fig2}c, ${\bf{r}}=(y,z)$, $H_{HSAI}(k_x,{\bf{r}})$ is the Hamiltonian for the HSAI and $G^r(E_F;k_x,{\bf{r}})$ is the retarded Green's function. 

\noindent The upper panel in Fig.~\ref{fig2}d presents the hinge current density $J_x(z)=\sum_{y=-L_y/2}^0j_x({\bf{r}})$ as a function of the layer index $z$. We see that $J_x(z)$ is verily confined on the hinge, in agreement with $\langle z/L_z\rangle$ and $A(k_x,E)$. Strikingly, this hinge current decays oscillatively into the side surface, exhibiting a beating mode (magenta line) in sharp contrast to that in spin-$1/2$ axion insulator~\cite{Gong2023Half}. This peculiar behavior can be quantitatively fitted by the superposition of two power-law decaying edge currents $J_x^1(z)=\frac{a_1}{\sqrt{z}}\cos{(2k_{F1}z+\phi_1)}$ and $J_x^2(z)=\frac{a_2}{\sqrt{z}}\cos{(2k_{F2}z+\phi_2)}$~\cite{Zou2022Half},  where $k_{F1}$ and $k_{F2}$ are the Fermi momenta for the two distinct modes marked by the white stars in Fig.~\ref{fig2}c, while $a_{1(2)}$ and $\phi_{1(2)}$ are fitting parameters. The integral of the current density over the layer index provides the current flux $I_x(z)=\int_{0}^zd\tilde{z}J_x(\tilde{z})$ (middle panel in Fig.~\ref{fig2}d), which oscillates around $2e/h$ and coincides perfectly with the fitting data. Additionally, the $z$-averaged current $\langle I_x(z)\rangle=\int_{0}^zd\tilde{z}I_x(\tilde{z})/z$ (red line) quantizes to $2e/h$ only a few layers away from the hinge. Imposing a finite thickness along $\hat{z}$-direction enables us to calculate the moving average current $\langle I_x(z)\rangle_{MA}=\int_{z-7}^{z+7}d\tilde{z}J_x(\tilde{z})$. The result displayed in the bottom panel demonstrates a helical hinge current quantized precisely to $\pm 2e/h$. Although the HSAI supports a QHHC identical to its integer surface Chern number, the topologically protected gapless exciations in lower dimension are completely absent as plotted in Fig.~\ref{fig2}b. It worths note that the energy gap in Fig.~\ref{fig2}b may be induced by the finite size effect. Our analysis in Supplementary Section 3 shows that the size dependence of the energy gap comes from the bulk bands, which demonstrates that this energy gap originates from the magnetic exchange interaction. Consequently, the conventional bulk-boundary correspondence that an integer Chern number must hold chiral edge states fails in HSAI, establishing an unusual quantum anomaly. 

\noindent\textbf{Non-reciprocal conductance}\\
\noindent Owing to the chirality bonded to the quantized surface Chern number, this QHHC can be unveiled by the non-reciprocal conductance $G^N_{ij}=G_{ij}-G_{ji}$ in a six-terminal device sketched in Fig.~\ref{fig2}e, where $G_{ij}$ is the differential conductance~\cite{datta_1995,Gong2023Half}. In this device, two longitudinal leads (terminals 5 and 6) are intimately connected to the two ends of the sample while four transverse leads (terminals 1, 2, 3, and 4) are attached to different hinges on the front surface. Figure.~\ref{fig2}f shows three representative non-reciprocal conductances versus the Fermi energy $E_F$ in the clean limit (solid lines), with non-magnetic Anderson disorder (dashed lines) and with magnetic Anderson disorder (dashed dotted lines). In general, $G^N_{65}=0$, $G^N_{31}=-2e^2/h$ and $G^N_{42}=2e^2/h$ when the Fermi level lies inside the band gap for all three cases, consistent with the current distribution in Fig.~\ref{fig2}d as well as the layer-resolved Chern numbers in Fig.~\ref{fig1}d. This verifies that the QHHC is topological protected as the quantized non-reciprocal conductance is immune to both non-magnetic and magnetic Anderson disorders that even breaks the global $\mathcal{PT}$ symmetry~\cite{Qi2011Topological,Hasan2010Colloquium}. 

\noindent 
Since multiple frequency ac current is robust against ambient perturbation, to detect this QHHC, we employ an alternate detection in which terminals 2, 4, 5, and 6 are grounded whereas a harmonic voltage $V(t)=V_0\sin{(\omega_0 t)}$ with a periodicity $T=2\pi/\omega_0$ is applied alternatively to terminal 1 or 3 as illustrated in Fig.~\ref{fig2}g. During the first (second) half period, a positive (negative) voltage is applied to terminal 1 (3) as an input while the current flows $i_3(t)$ [$i_1(t)$] from terminal 3 (1) is detected as an output. Their combination gives rise to an asymmetric net current $i(t)=i_1(t)+i_3(t)$ as shown in Fig.~\ref{fig2}h. Performing a Fourier transform converts $i(t)$ into $I(\omega)$. The non-reciprocal conductance can then be determined from the equation $F(\omega)=|I(\omega)(\omega^2-\omega_0^2)|/(2N\omega_0V_0)$ with $N$ the number of periodicity (see Supplementary Section 4 for details). The result displayed in Fig.~\ref{fig2}i shows that $F(\omega)=G^N_{13}$ when $\omega=2\omega_0$.
Thus, non-reciprocal conductance measurements offer a reliable experimental method to visualize the QHHC in HSAI. 
   
\noindent\textbf{Axion term}\\
\noindent The HSAI can alternatively be characterized by the axion term~\cite{Qi2008Topological}, which can be evaluated directly from the hybrid Wannier functions (HWFs) constructed in terms of the Bloch wavefunctions~\cite{Varnava2020Axion}. In this scenario, the axion term on a slab is defined as~\cite{Varnava2020Axion}
\begin{align}\label{the}
\theta_{CS}^{slab}=-\frac{1}{L_z}\int{d^2{\bf{k}}}\sum_n[z_{n{\bf{k}}}\tilde{\Omega}_{{\bf{k}}nn}^{xy}],
\end{align}
where $z_{n{\bf{k}}}$ is the hybrid Wannier charge center along $\hat{z}$-direction and $\tilde{\Omega}_{{\bf{k}}nn}^{xy}$ is corresponding non-Abelian Berry curvature. 

\begin{figure}[t]
  	\centering
    	\includegraphics[width=0.9\textwidth]{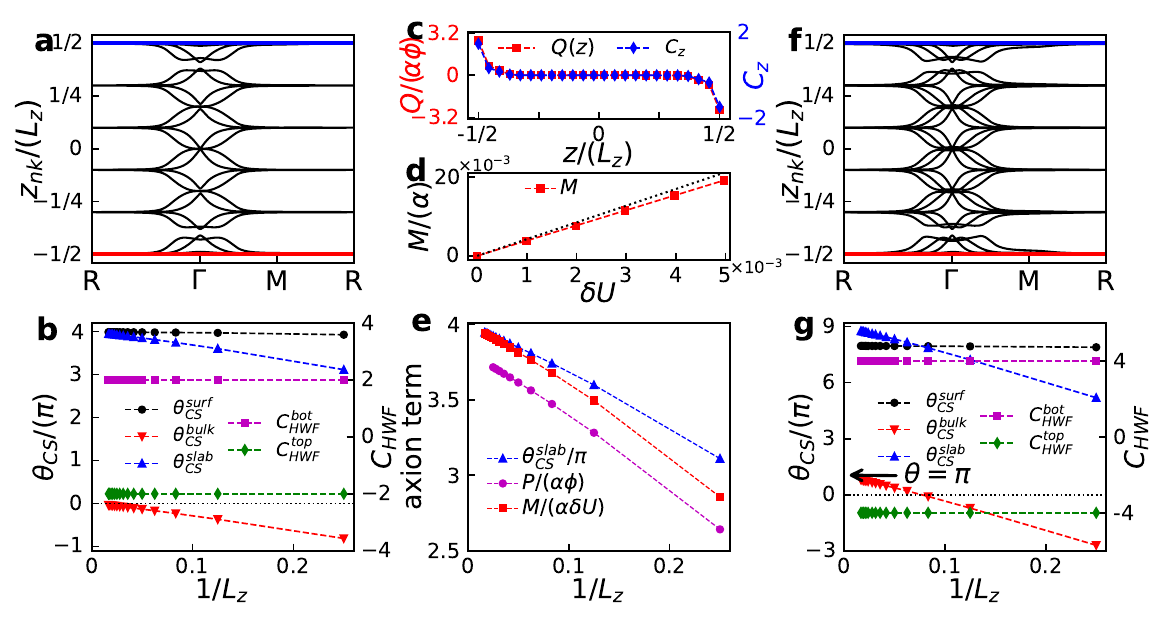}
    	\caption{\textbf{Axion term and topological magneto-electric effect.}
	\textbf{a} and \textbf{f} Hybrid Wannier charge centers  $z_{n{\bf{k}}}$ along R$\rightarrow$$\Gamma$$\rightarrow$M$\rightarrow$R loop inside the first Brillouin zone for a six-layer HSAI slab with spin-$3/2$ (\textbf{a}) and spin-$5/2$ (\textbf{f}), respectively.
	\textbf{b} and \textbf{g}  are corresponding axion terms and the surface Chern numbers versus the inverse layer thickness obtained by using the HWFs.
	\textbf{c} Magnetic field induced charge distribution along $\hat{z}$-direction and the layer-resolved Chern number for a spin-$3/2$ HSAI with $L_z=24$. Here, the charge polarization is obtained on a HSAI slab with open boundary condition along $\hat{y}$-direction ($L_y=40$) but periodic boundary condition along $\hat{x}$-direction. The magnetic flux inside one unit cell is $\phi_0=Ba_0^2=0.01h/e$. 
	\textbf{d} Electric field induced orbital magnetization for a spin-$3/2$ HSAI with $L_z=20$. The black dashed line shows the ideal case (IC) with an exact axion term $\theta=4\pi$.
	\textbf{e} Size scaling of the axion term $\theta_{CS}^{slab}/\pi$, polarization coefficient $P/(\alpha\phi)$, and magnetization coefficient $M/(\alpha\delta U)$ at $\delta U=0.001$. We have checked that the slight deviation of $P/(\alpha\phi)$ originates from the finite size effect.
		} 
    	\label{fig3}
    	%\vspace{80pt}
\end{figure}

\noindent Figure~\ref{fig3}(a) shows $z_{n{\bf{k}}}$ in the first Brillouin zone for a six-layer HSAI with spin-$3/2$.
These $z_{n{\bf{k}}}$ consist of two different types, those localized on the top and bottom surfaces as emphasized by the red and blue lines and those extending into the bulk denoted by black lines. Those surface Wannier bands will disappear under a periodic boundary condition when connecting the top and bottom surfaces. The total axion term of the slab can subsequently be divided into two parts $\theta_{CS}^{slab}=\theta_{CS}^{bulk}+\theta_{CS}^{surf}$ with $\theta_{CS}^{bulk}$ and $\theta_{CS}^{surf}$ the axion terms corresponding to the surface and bulk HWFs. The bulk axion term $\theta_{CS}^{bulk}$ is identical to that obtained analytically from the Chern-Simons three form in the infinite size limit\cite{Varnava2020Axion}. In Fig.~\ref{fig3}b, we show $\theta_{CS}^{bulk}$ (red upside down triangle), $\theta_{CS}^{surf}$ (black circle) together with $\theta_{CS}^{slab}$ (blue triangle) versus the inverse thickness $1/L_z$. There are three distinctive features in this figure. First, the total axion term shows an obvious tendency quantized to $\theta_{CS}^{slab}=4\pi$ when the system size approaches infinity ($1/L_z\rightarrow 0$), which confirms the quadruple axion term in two dimensional HSAI slab. Second, the axion term originates completely from the surface HWFs although the bulk HWFs also result in a small value that decreases $\theta_{CS}^{bulk}$ at finite size. Third, the axion term $\theta_{CS}^{surf}$ obeys the relation $\theta_{CS}^{surf}=(C_{HWF}^{bot}-C_{HWF}^{top})\pi$ in the infinite size limit, where $C_{HWF}^{top(bot)}$ is the top (bottom) surface Chern number obtained from the surface HWFs as indicated by cyan diamond (magenta square). These peculiar results reaffirm the unusual quantum anomaly and also the quadruple axion term $\theta_{HSAI}=4\pi$ in HSAI with spin-$3/2$.
   
\noindent\textbf{Topological magneto-electric effect}\\
\noindent Such quadruple axion term implies a unique topological magneto-electric effect~\cite{Wang2022Topological,Li2022Identifying}. When applying a magnetic field $B_z$ to the HSAI along $\hat{z}$-direction, the Hamiltonian in Eq.~\ref{model_Ham} acquires a Peierls phase~\cite{Li2022Identifying}, which redistributes the electron charge $Q(z)$ in accordance to the confined  layer-resolved Chern number $C_z$ as shown in Fig.~\ref{fig3}c. The ensuing charge polarization $P=\sum_{z=-L_z/2}^{L_z/2}zQ(z)/L_z$ almost quantizes to $P\approx 4\alpha\phi$, where $\alpha$ is the fine structure constant and $\phi$ is the total magnetic flux penetrating the HSAI slab. In comparison, a quantized orbital magnetization can emerge under an external electric field $E_z$ when incurring a potential drop $\delta U=eE_zL_z$ in the HSAI Hamiltonian~\cite{Thonhauser2005Orbital}. The red square in Fig.~\ref{fig3}(d) shows the orbital magnetization $M$ as a function of $\delta U$, which agrees quantitatively well with the ideal case benchmarked by the black line. These two results independently certify the quadruple axion
term $\theta_{HSAI}$ in spin-$3/2$ HSAI. The slight deviation from the exactly quadruple value originates from the finite size effect, which is further revealed by the size scalings of the axion term $\theta_{CS}^{slab}/\pi$, polarization coefficient $P/(\alpha\phi)$, and magnetization coefficient $M/(\alpha\delta U)$ against the inverse layer thickness $1/L_z$ in Fig.~\ref{fig3}e. 
We also evaluate the axion term and the surface Chern numbers for spin-$5/2$ HSAI in terms of the HWFs (Fig.~\ref{fig3}f). The results displayed in Fig.~\ref{fig3}g demonstrate that spin-$5/2$ HSAI possesses a surface Chern number $C^{top(bot)}_{HWF}=\mp 4$, a total axion term $\theta^{slab}_{CS}=9\pi$, a surface axion term $\theta^{surf}_{CS}=8\pi$ and a bulk axion term $\theta^{bulk}_{CS}=\pi$. Systematic results for the spin-$5/2$ HSAI are provided in Supplementary Section 5. The topological properties for HSAI with different spin species are summarized in Table.~\ref{table1}, giving a distinct axion field $\theta=(s+1/2)^2\pi$ and $C_{surf}^{top/bot}=\mp(1/2+3/2+\cdots+s)$.

\begin{figure}[t]
  	\centering
    	\includegraphics[width=0.9\textwidth]{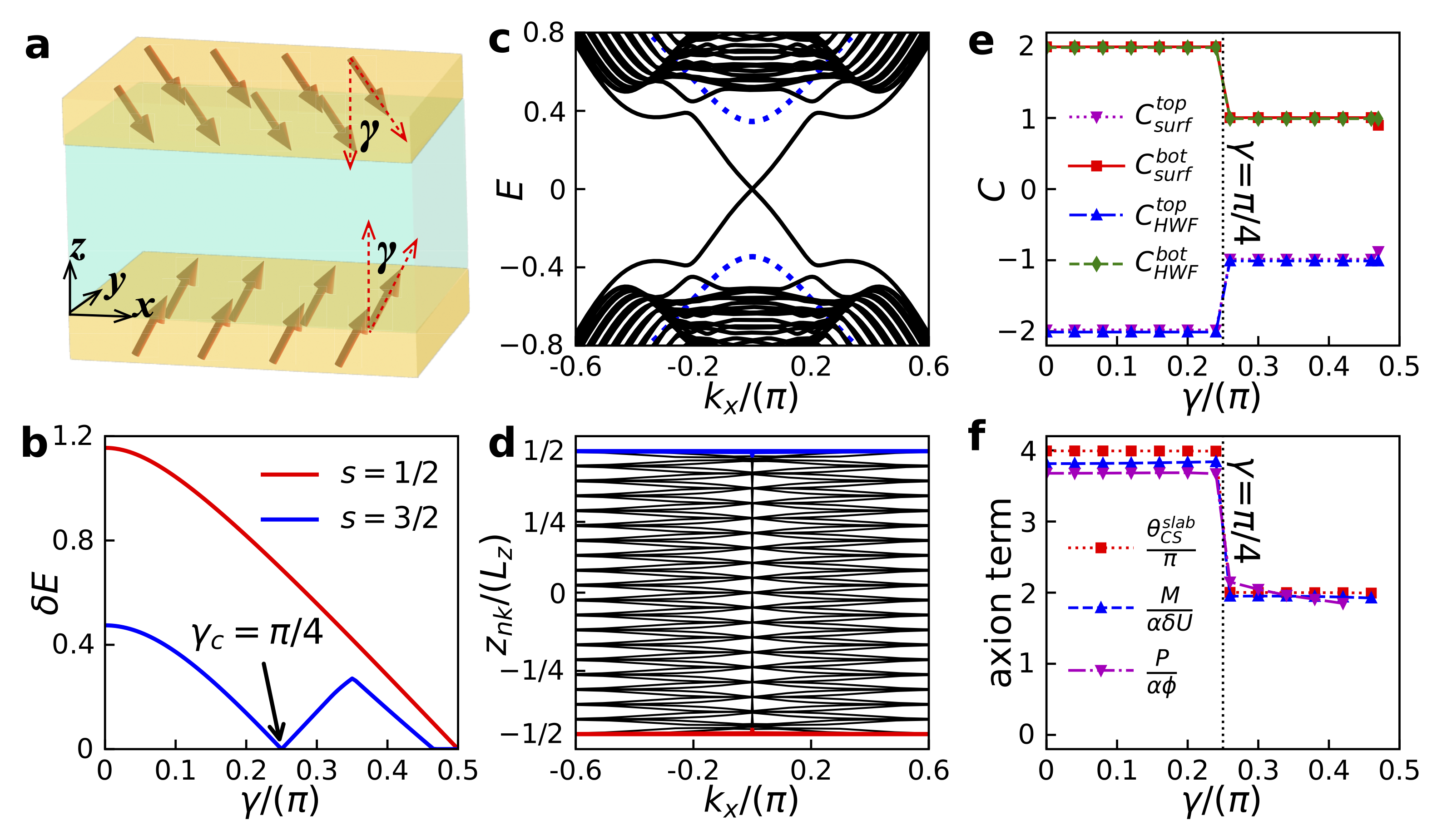}
    	\caption{\textbf{Phase transition in spin-$3/2$ HSAI.}
	\textbf{a} Canted HSAI under an in-plane magnetic field. $\gamma$ is the canting angle between the magnetic vector and $\hat{z}$-axis (polar angle).
	\textbf{b} Energy gaps versus $\gamma$ for spin-$3/2$ HSAI and for spin-$1/2$ axion insulator, respectively.
	\textbf{c} Energy spectra for the HSAI with spin-$3/2$ (black solid lines) and spin-$1/2$ (blue dashed lines) at $\gamma=\pi/4$.
	\textbf{d} Hybrid Wannier charge center $z_{n{\bf{k}}}$ as a function of $k_x$ for a HSAI slab at $\gamma=\pi/4$.
	\textbf{e} Surface Chern numbers obtained from the effective Hamiltonian in Eq.~(\ref{model_Ham}) and the HWFs versus $\gamma$.
	\textbf{f} Axion term $\theta_{CS}^{slab}/\pi$, polarization coefficient $P/(\alpha\phi)$ ($\phi_0=0.01h/e$) and magnetization coefficient $M/(\alpha\delta U)$ ($\delta U=0.001$) versus  $\gamma$.
	The thickness of the HSAI is $L_z=20$.
		} 
    	\label{fig4}
    	%\vspace{80pt}
\end{figure}

\noindent\textbf{Tunable topological phase transition}\\
\noindent In the presence of an in-plane magnetic field, the antiparallel magnetic moments in the top and bottom layers become canted with the canting angle $\gamma$ proportional to the magnetic field strength as illustrated in Fig.~\ref{fig4}a. In this case, the quantized axion field in the infinite size limit is protected by $m_x\mathcal{P}$ symmetry where $m_x$ is the mirror plane normal to $x$-direction. In Fig.~\ref{fig4}b, we compare two dimensional band gaps as functions of $\gamma$ for spin-$1/2$ axion insulator and spin-$3/2$ HSAI. Because the exchange gap is determined by the perpendicular magnetization $M_z$, the band gap for spin-$1/2$ axion insulator decreases monotonically as $\gamma$ is enlarged and finally becomes zero when $\gamma=\pi/2$. On the contrary, the band for spin-$3/2$ HSAI exhibits a gap close at $\gamma=\pi/4$ as shown in Fig.~\ref{fig4}c, suggesting a possible topological phase transition. Indeed, Figure~\ref{fig4}e shows that the surface Chern number obtained using both the Bloch wavefunctions and the HWFs(Fig.~\ref{fig4}d) changes from $+2$ ($-2$) to $+1$ ($-1$) when $\gamma=\pi/4$. At this point, the HWFs are connected at the $\Gamma$ point (Fig.~\ref{fig4}d), therefore the Berry curvature and the surface Chern number can transfer from one side to the other~\cite{Olsen2017Surface}, leading to an axionic phase transition.
Such topological phase transition is further affirmed by the axion field, the polarization and magnetization coefficients shown in Fig.~\ref{fig4}f.

\noindent\textbf{Discussion}\\
\noindent The device application of axion insulators requires the fine-control of the transport signals such as the magneto-electric response or the QHHC, which are identical to the axion field. In spin-$1/2$ axion insulators, the axion term cannot be tuned  without disrupting the existing $\mathcal{S}$ symmetry or refabricating the setup~\cite{Coh2011Chern}. Nevertheless, because different surface bands shown in Fig.~\ref{fig1}b can be coupled via the in-plane exchange interaction ($M_xs_x\otimes\tau_0$), an apparent topological phase transition between axion insulators with different axion fields occurs in the HSAI. Consequently, the axion term $\theta_{HSAI}$ (in unit of $\pi$), hence the QHHC $G^N_{ij}$ (in unit of $e^2/2h$) and the magneto-electric effect $P/(\alpha B)$ [$M/(\alpha E)$]  in HSAI can be precisely adjusted from $4$ to $2$ via the application of an external magnetic field.  
Thus, our work opens up an exciting possibility for the groundbreaking advancement in the practical application of axion insulators.

\noindent In conclusion, we have proposed a HSAI defined on the high spin space and shown that this HSAI possesses a multiple axion field protected by the combined lattice and time-reversal symmetry. Notably, the axion term in the bulk of a HSAI still quantizes to $\theta=0$ or $\theta=\pi$ while the surface of HSAI possesses a large axion term and a consistent integer Chern number, which can be tuned by manipulating the magnetic configuration through an external magnetic field. 
These results extend the scope of recently discovered axion insulator in magnetic topological materials. 
In ultra-cold fermions on a honeycomb lattice, the exchange gap can be introduced by complex next-nearest-neighbour tunneling terms through circular modulation of the lattice position~\cite{Jotzu2014Experimental}.  We thus propose that our theory can be tested in high spin ultra-cold fermions on a stacked honeycomb lattice, where the non-reciprocal conductance can be detected by the orthogonal drifts analogous to a Hall current under a constant force to the atoms~\cite{Jotzu2014Experimental,Jean2012Conduction}.\\

\noindent\textbf{Methods}\\
\noindent \textbf{Caltulations of the layer-resolved Chern number, magnetization, and polarization.} In a HSAI slab with periodic boundary conditions along the lateral dimensions, the momenta $k_x$ and $k_y$ are good quantum numbers because of the translation symmetry. Therefore, the layer-resolved Chern number can be calculated by projecting the total Chern number into specific layer, which can be written as
\begin{align}\label{layer_cnum}
C_z=\frac{1}{\pi}\sum_{E_m({\bf{k}})<E_F<E_n({\bf{k}})}\int{dk_xdk_y}\text{Im}\frac{\langle m_{\bf{k}}|\hat{P}_z\partial_{k_x}H_{HSAI}|n_{\bf{k}}\rangle\langle n_{\bf{k}}|\partial_{k_y}H_{HSAI}|m_{\bf{k}}\rangle}{[E_m({\bf{k}})-E_n({\bf{k}})]^2}.
\end{align}
Here, $E_F$ is the Fermi energy, $E_{m(n)}({\bf{k}})$ is the eigenenergy of $H_{HSAI}$ with $|m_{\bf{k}}\rangle$ ($|n_{\bf{k}}\rangle$) the corresponding eigenstates, $\hat{P}_z=|\psi_z\rangle\langle\psi_z|$ is the projecting operator. The integral is performed inside the first Brillouin zone. 

\noindent Under an electric field $E_z$ along $\hat{z}$-direction, a potential drop occurs inside the HSAI slab along the same direction. The onsite energy in each layer acquires an additional value $eE_zz$ with $z$ the layer index and the total potential drop in the HSAI slab is $\delta U=eE_zL_z$. The orbital magnetization can then be obtained accordingly by using
\begin{align}\label{mag}
M=\frac{-e}{2\pi h}\sum_{\tilde{E}_m<E_F<\tilde{E}_n}\int{dk_xdk_y}\text{Im}\frac{(\tilde{E}_m+\tilde{E}_n-2E_F)}{(\tilde{E}_m-\tilde{E}_n)^2}\langle \tilde{m}|\partial_{k_x}\tilde{H}_{HSAI}|\tilde{n}\rangle\langle \tilde{n}|\partial_{k_y}\tilde{H}_{HSAI}|\tilde{m}\rangle,
\end{align}
where $\tilde{H}_{HSAI}=H_{HSAI}+eE_zz$ with $\tilde{E}_{m(n)}$ and $|\tilde{m}\rangle$ ($|\tilde{n}\rangle$) its eigenenergy and eigenstate, respectively.

\noindent Applying a magnetic field $B_z$ along $\hat{z}$-direction introduces a gauge potential to the HSAI lattice and thus breaks the in-plane translation symmetry. Inside each unit cell, HSAI acquires a gauge field $\phi_0=\int{d{\bf{r}}}{\bf{A}}\cdot{\bf{r}}/\Psi_0$ with $\Psi_0=h/(2e)$ the magnetic flux quantum. The total magnetic flux penetrating the HSAI slab is $\phi=B_zL_xL_y$. We adopt the Landau gauge ${\bf{A}}=\begin{pmatrix}-yB_z,&0,&0\end{pmatrix}$, so the translation symmetry along $\hat{y}$-direction is broken while that along $\hat{x}$-direction sustains. In this case, the charge distribution induced by the magnetic field can be obtained by using the Green's function method, yielding
\begin{align}\label{polarization}
q(z)=\frac{e}{\pi}\sum_{x,y}\int_{-\infty}^{E_F}{dE}\text{Im}\text{Tr} G^r(E,{\bf{r}}),
\end{align}
where ${\bf{r}}=\begin{pmatrix}x,&y,&z\end{pmatrix}$ and the Green's function $G^r(E,{\bf{r}})=(E+i\eta-H_{HSAI})^{-1}$ with $\eta$ the imaginary line width function. On the other hand, as $k_x$ is still a good quantum number, the charge distribution along $\hat{z}$-direction can be alternatively obtained by using 
\begin{align}\label{k_polarization}
q(z)=\frac{e}{2\pi^2}\sum_{y}\int_{-\infty}^{E_F}{dE}\int{dk_x}\text{Im}\text{Tr} G^r(E,k_x,y,z).
\end{align}
Moreover, because only the negative charge originating from electrons are considered here in Eqs.~\ref{polarization} and ~\ref{k_polarization}, to derive the unbalanced charge distribution and in turn the polarization, the uniform background charge compensating the positive ions in the lattice has to be removed from the results, which has the form $q_{back}=-\sum_{z=-L_z/2}^{L_z/2}q(z)/L_z$ because of the charge conservation. As a result, the charge distribution has the form $Q(z)=q(z)+q_{back}$. The charge polarization can finally be expressed as $P=\sum_{z=-L_z/2}^{L_z/2}zQ(z)/L_z$.

\noindent \textbf{Caltulations of the axion term using the hybrid Wannier function.} In a HSAI slab, the hybrid Wannier wavefunction $|h_{n,{\bf{k}}}\rangle$ can be constructed from the Bloch wave function. We thus have $|h_{n,{\bf{k}}}\rangle=1/2\pi\int_{-\pi}^{\pi}dk_z|n_{\bf{k}}\rangle e^{-i({\bf{k}}\cdot{\bf{r}}+k_zz)}$. In this case, the hybrid Wannier charge center takes the form $z_{n_{\bf{k}}}=\langle h_{n,{\bf{k}}}|z|h_{n,{\bf{k}}}\rangle$~\cite{Varnava2020Axion}. To calculate the non-Abelian Berry curvature, we divide the two-dimensional Brillouin zone into a regular mesh with $b_x$ and $b_y$ being the primitive reciprocal vectors that define the mesh. Then the gauge covariant Berry curvature has the form~\cite{Ceresoli2006Orbital} 
\begin{align}\label{gauge_covariant_Berry_curvature}
\tilde{\Omega}_{{\bf{k}}nn}^{xy}=i(\langle\tilde{\partial}_xh_{n,{\bf{k}}}|\tilde{\partial}_yh_{n,{\bf{k}}}\rangle-\text{h.c.}),
\end{align}
where $|\tilde{\partial}_ih_{n,{\bf{k}}}\rangle=(|\tilde{h}_{n,{\bf{k}}+b_i}\rangle-|\tilde{h}_{n,{\bf{k}}-b_i}\rangle)/2$. The wavefunctions constructed by a linear combination of the occupied bands at neighboring mesh point are $|\tilde{h}_{n,{\bf{k}}\pm b_i}\rangle=\sum_{n^{\prime}}(S_{{\bf{k}},{\bf{k}}\pm b_i}^{nn\prime})^{-1}$ $\times|h_{n^{\prime},{\bf{k}}\pm b_i}\rangle$, where the matrix $S_{{\bf{k}},{\bf{k}}^\prime}^{nn\prime}=\langle h_{n,{\bf{k}}}|h_{n^{\prime},{\bf{k}}^\prime}\rangle$.

\noindent \textbf{Green's function method for calculating the differential conductance $G_{ij}$}. The differential conductance $G_{ij}$ corresponds to the transmission coefficient $T_{ij}$ from terminal $j$ to terminal $i$, which can be derived by using the non-equilibrium Green's function method. Based on the Landauer-B\"uttiker formula~\cite{datta_1995}, the transmission coefficient $T_{ij}$ can be expressed as $T_{ij}=\text{Tr}[\Gamma_{i}G^r\Gamma_{j}G^a]$, where $\Gamma_{i(j)}=i[\Sigma_{i(j)}-\Sigma_{i(j)}^{\dagger}]$ is the line width function and $G^{r}=(G^a)^\dagger=[E_F+i\eta-H_{HSAI}-\sum_i\Sigma_i]^{-1}$ with $E_F$ the Fermi energy, $\eta$ the imaginary line width function and $\Sigma_{i}$ the self energy due to the coupling to terminal $i$.  To incorporate the disorders, we generate random potentials $\delta E\in \begin{pmatrix}-W/2,&W/2\end{pmatrix}$ for the non-magnetic Anderson disorders or $\delta M_z\in \begin{pmatrix}-W_z/2,&W_z/2\end{pmatrix}$ for magnetic Anderson disorders at each site ${\bf{r}}$, then add these random potentials to the Hamiltonian in the Green's functions. The results in the presence of disorders are calculated under $10$ times average (Fig.~\ref{fig2}f).

\noindent\textbf{Data availability}\\
  	The data that support the plots within this paper and other findings of this study are available from the corresponding author upon request. Source data are provided with this paper.\\

    \noindent\textbf{Code availability}\\
  	The code that is deemed central to the conclusions is available from the corresponding author upon request.\\
\\
    \noindent\textbf{References}
    \bibliographystyle{naturemag}
    \bibliography{ref}
    \noindent\textbf{Acknowledgements}\\
    We are grateful for the fruitful discussions with Dr. Zhiqiang Zhang and Prof. Chui-Zhen Chen. Y.-H.L acknowledges the support from the Fundamental Research Funds for the Central Universities. H.J. acknowledges the support from the National Key R\&D Program of China (Grants No. 2019YFA0308403 and No. 2022YFA1403700) and the National Natural Science Foundation of China (Grant No. 12350401). X.C.X. acknowledges the support from the Innovation Program for Quantum Science and Technology (Grant No. 2021ZD0302400)
    \\ 
    
    \noindent\textbf{Author contributions}\\
    H.J. conceived the idea of high spin axion insulators after a discussion with Y.-H.L and X.C.X.. S.L. performed calculations with assistance from M.G., Y.-H.L. and H.J.. S.L. and Y.-H.L. wrote the manuscript with contributions from all authors. H.J. and X.C.X. supervised the project.\\
\\  
    \noindent\textbf{Competing interests}\\
    The authors declare no competing interests.\\
    
    \noindent\textbf{Additional information}\\ 
    \textbf{Supplementary information} The online version contains supplementary material available at .\\
    \textbf{Correspondence} and requests for materials should be addressed to Y.-H. L. or H. J.\\

    \setcounter{figure}{0}
    \captionsetup[figure]{labelfont={bf}, name={Extended Data Fig.}, labelsep=period}

\end{document}

\maketitle
	\tableofcontents
    
    \newpage

{\section{Spin matrices for different spins}\label{sec_spin_mat}}
The spin matrices for different spin species can be generated mathematically from the Clifford algebra~\cite{ Lounesto,Lachieze}, which are

\begin{table}[ht]
\begin{tabular}{ |p{1cm}||p{4.2cm}|p{6.3cm}|p{3.7cm}|  }
 \hline
 spin-$s$& $s_x$ & $s_y$& $s_z$ \\
 \hline
1/2&$\begin{pmatrix}0,&1\\1,&0\end{pmatrix}$    &$\begin{pmatrix}0,&-i\\i,&0\end{pmatrix}$         &$\begin{pmatrix}1,&0\\0,&-1\end{pmatrix}$\\
\hline
3/2&$\begingroup
\setlength\arraycolsep{2pt}\begin{pmatrix}
		0&\frac{\sqrt{3}}{2}&0&0\\
		\frac{\sqrt{3}}{2}&0&1&0\\
		0&1&0&\frac{\sqrt{3}}{2}\\
		0&0&\frac{\sqrt{3}}{2}&0
	\end{pmatrix}\endgroup$  &$\begingroup
\setlength\arraycolsep{2pt}\begin{pmatrix}
     	0&-i\frac{\sqrt{3}}{2}&0&0\\
     	i\frac{\sqrt{3}}{2}&0&-i&0\\
     	0&i&0&-i\frac{\sqrt{3}}{2}\\
     	0&0&i\frac{\sqrt{3}}{2}&0
     \end{pmatrix}\endgroup$ &$\begingroup
\setlength\arraycolsep{2pt}\begin{pmatrix}
     	\frac{3}{2}&0&0&0\\
     	0&\frac{1}{2}&0&0\\
     	0&0&-\frac{1}{2}&0\\
     	0&0&0&-\frac{3}{2}
     \end{pmatrix}\endgroup$      \\
\hline
5/2&$\begingroup
\setlength\arraycolsep{1pt}
\begin{pmatrix}
0&\frac{\sqrt{5}}{2}&0&0&0&0\\
		\frac{\sqrt{5}}{2}&0&\sqrt{2}&0&0&0\\
		0&\sqrt{2}&0&\frac{3}{2}&0&0\\
		0&0&\frac{3}{2}&0&\sqrt{2}&0\\
		0&0&0&\sqrt{2}&0&\frac{\sqrt{5}}{2}\\
		0&0&0&0&\frac{\sqrt{5}}{2}&0
\end{pmatrix}
\endgroup$  &$\begingroup
\setlength\arraycolsep{1pt}\begin{pmatrix}
		0&-i\frac{\sqrt{5}}{2}&0&0&0&0\\
		i\frac{\sqrt{5}}{2}&0&-i\sqrt{2}&0&0&0\\
		0&i\sqrt{2}&0&-i\frac{3}{2}&0&0\\
		0&0&i\frac{3}{2}&0&-i\sqrt{2}&0\\
		0&0&0&i\sqrt{2}&0&-i\frac{\sqrt{5}}{2}\\
		0&0&0&0&i\frac{\sqrt{5}}{2}&0
	\end{pmatrix}\endgroup$ &$\begingroup
\setlength\arraycolsep{2pt}\begin{pmatrix}
		\frac{5}{2}&0&0&0&0&0\\
		0&\frac{3}{2}&0&0&0&0\\
		0&0&\frac{1}{2}&0&0&0\\
		0&0&0&-\frac{1}{2}&0&0\\
		0&0&0&0&-\frac{3}{2}&0\\
		0&0&0&0&0&-\frac{5}{2}
	\end{pmatrix}\endgroup$\\
 \hline
\end{tabular} 
 \end{table}

{\section{Hamiltonian for high spin axion insulator from symmetry perspective}\label{Ham}}

The model Hamiltonian for the high spin axion insulator can be built from the high spin topological insulator preserving both parity and time-reversal symmetry~\cite{Li2010Dynamical}.
These parity and time-reversal symmetry on the orbital and spin basis are defined as $\mathcal{P}=\tau_z$ and $\mathcal{T}=e^{-is_y}\mathcal{K}$, respectively, where $\tau_z$ is the Pauli matrix acting on orbitals, $s_y$ is the spin matrix shown in Sec.~\ref{sec_spin_mat}, and $\mathcal{K}$ is the complex conjugation operator. Since (the odd power of) the spin matrices $s_{x,y,z}$ flip sign under the time-reversal symmetry, the minimum Hamiltonian for a high-spin topological insulator preserving both parity and time-reversal symmetry thus reads
\begin{align}\label{TI_Ham}
H_0=(m_0-Bk^2)s_0\otimes\tau_z+\sum_{i=x,y,z}A_ik_is_i\otimes\tau_x,
\end{align}
where $k^2=k_x^2+k_y^2+k_z^2$, $m_0$, $B$, $A_{i=x,y,z}$ are system parameters. The first term in Supplementary Eq.~(\ref{TI_Ham}) is the kinetic energy while the second term represents the spin-orbital coupling. Due to the time-reversal symmetry, the axion field of $H_0$ is quantized. In the main text, we take the spin orbit coupling $A_x=A_y=A_z$ for convenience. However, our theory is universal and is not limited to special parameters. It also worths note that the spin orbital coupling is crucial for the quantized axion field, which otherwise becomes $\theta_0=0$ if $A_{i=x,y,z}=0$. In the presence of magnetic ordering, the time-reversal symmetry is explicitly broken. However, in the antiferromagnetic phase, the 
combined lattice and time-reversal symmetry is still well preserved because the magnetic moments on the adjacent layers are antiparallel. The magnetic layers introduces an exchange interaction into the system, which recasts the Hamiltonian as
\begin{align}\label{sys_Ham}
H=H_0+\Delta{\bf{m}}_s\cdot{\bf{s}}\otimes\tau_0,
\end{align}
where $\Delta$ is the exchange gap between the magnetic moment ${\bf{m}}_s$ and the topological electrons with spin ${\bf{s}}$. In three dimension limit, the system preserves the symmetry $\mathcal{S}=T\tau_{1/2}$ with  $\tau_{1/2}$ the half translation operator along $z$-direction. Whereas, this $\mathcal{S}$ symmetry breaks into combined parity and time-reversal symmetry that can be defined as $\mathcal{PT}=\sigma_ze^{-is_y}\mathcal{K}$ on a slab geometry, where $\sigma_z$ is the Pauli matrix switching the magnetic moments between the top and bottom layers

{\section{Finite size effect}\label{scaling_sec}}

\begin{figure*}[ht]
    \centering
    \includegraphics[width=0.62\linewidth]{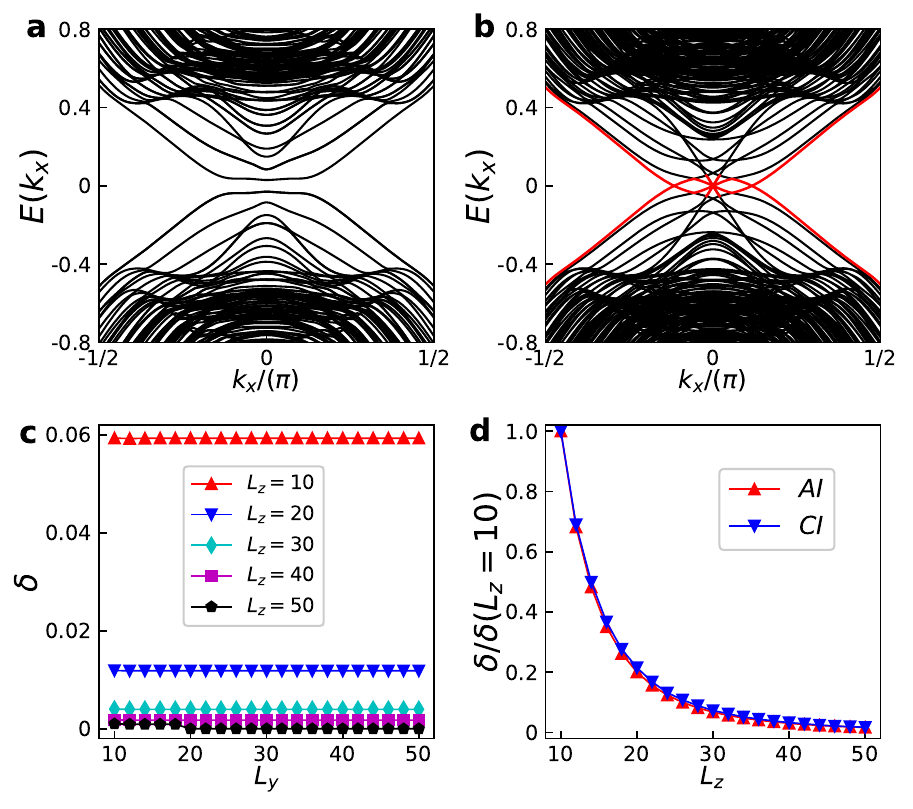}
    \caption{\textbf{Size dependence of the energy gap.} \textbf{a} and \textbf{b} show the band structures for systems in the antiferromagnetic and ferromagnetic cases, respectively. In the ferromagnetic case, the system is a spin-3/2 Chern insulator. The red lines in \textbf{b} denote the four gapless edge states. Here, the system size is $L_y=20$, $L_z=10$. \textbf{c}, energy gap for the HSAI versus $L_y$. \textbf{d}, size dependence of the HSAI band gap (red triangle) and Chern insulator bulk band gap (blue up-side-down triangle) with $L_y=20$. The results for the Chern insulator bulk band gap is obtained after removing the surface states (red lines). All parameters for the model Hamiltonian are exactly the same as those in the main text.
	}
    \label{siz_scal}
\end{figure*}

To rule out the possibility of the finite-size effect induced energy gap in HSAI, we study the size dependence of the energy gap. Supplementary figures~\ref{siz_scal} a and b show the band structures of a one-dimensional nanowire with size $L_y=20$ and $L_z=10$ for the system in the antiferromagnetic and ferromagnetic cases, respectively. In the ferromagnetic case, the system is a spin-3/2 Chern insulator with a Chern number $C=4$ (see Sec.~\ref{even_odd}). Therefore, there are four gapless edge states denoted by the red lines in Supplementary Fig.~\ref{siz_scal} b, in sharp contrast to the gapped band structure for a HSAI in Supplementary Fig.~\ref{siz_scal} a. On the other hand, figure~\ref{siz_scal} c shows the band gap for the HSAI as a function of $L_y$. It is apparent that this gap is independent of the size $L_y$. To further rule out the possibility of the finite-size effect along $z$-direction, we compare the band gap for the HSAI with the bulk band gap of the Chern insulator after removing the edge bands (red lines in Supplementary Fig.~\ref{siz_scal}b). The results plotted in Supplementary Fig.~\ref{siz_scal}d show that the two band gaps as functions of $L_z$ coincide quantitatively with each other, which demonstrates that the band gap in HSAI is also induced by the bulk bands rather than the finite size effect due to the overlapping between edge states on top and bottom surfaces.

{\section{Alternate detection}\label{alter_detect}}
The temporal dependent asymmetric current output in the main text has the form
\begin{align}\label{t_curr}
i(t)=\left\{\begin{array}{rcl}V_0G_{31}\sin{\omega_0t},&&{{2n\pi}/{\omega_0}<t\le{(2n+1)\pi}/{\omega_0}}\\
V_0G_{13}\sin{\omega_0t},&&{{(2n+1)\pi}/{\omega_0}<t\le {(2n+2)\pi}/{\omega_0}}
\end{array}\right.
\end{align}
for an arbitary integer $n$, where $G_{ij}$ is the conductance from terminal $j$ to terminal $i$ as explained in the main text, $V_0$ and $\omega_0$ are the amplitude and frequency of the alternative harmonic voltage input, respectively.
Performing a Fourier transform with respect to $t$ converts the current into the frequency domain, which yields
\begin{align}\label{ome_curr}
\begin{split}
I(\omega)&=\frac{\omega_0}{2\pi}\int_{-\pi/\omega_0}^{\pi/\omega_0}dte^{-i\omega t}i(t)\\
&=\sum_{n=-N/2}^{N/2}V_0\left[G_{31}\int_{2n\pi/\omega_0}^{(2n+1)\pi/\omega_0}dte^{-i\omega t}\sin{\omega_0t}+G_{13}\int_{(2n+1)\pi/\omega_0}^{(2n+2)\pi/\omega_0}dte^{-i\omega t}\sin{\omega_0t}\right]\\
&=\sum_{n=-N/2}^{N/2}\frac{V_0\omega_0e^{-i2n\pi\omega/\omega_0}}{\omega_0^2-\omega^2}(1+e^{-i\pi\omega/\omega_0})(G_{13}e^{-i\pi\omega/\omega_0}-G_{31}).
\end{split}
\end{align}
Therefore, the non-reciprocal conductance can be unveiled by the function $F(\omega)=|I(\omega)(\omega^2-\omega_0^2)|/(2N\omega_0V_0)$ with $N$ the truncation of the summation $n$. If $\omega=2\omega_0$, $F(\omega)=G_{13}-G_{31}=G^N_{13}$ with $G^N_{13}$ the non-reciprocal conductance when $N$ approaches infinity. However, the function $F(\omega)=0$ if otherwise. Consequently, the quantized helical hinge current can be readily detected by using the alternate method proposed in the main text.

{\section{High spin axion insulator with spin-$\frac{5}{2}$}}
The spin matrices for spin-$5/2$ are $6\times6$ as shown in Sec.~\ref{sec_spin_mat} because the magnetic quantum number $m_z$ has six values ranging from $-5/2$ to $5/2$. As illustrated in Supplementary Fig.~\ref{sfig1}a, the model Hamiltonian for the high spin axion insulator (HSAI) defined on spin-${5}/{2}$ space shares the same form as the spin-${3}/{2}$ HSAI despite that the spin matrices are different. Thus, all quantities presented in the main text can be obtained identically by incorporating the spin matrices for spin-${5}/{2}$. It is important to note that all parameters in this supplementary information are the same as those in the main text.

\begin{figure*}[ht]
    \centering
    \includegraphics[width=0.6\linewidth]{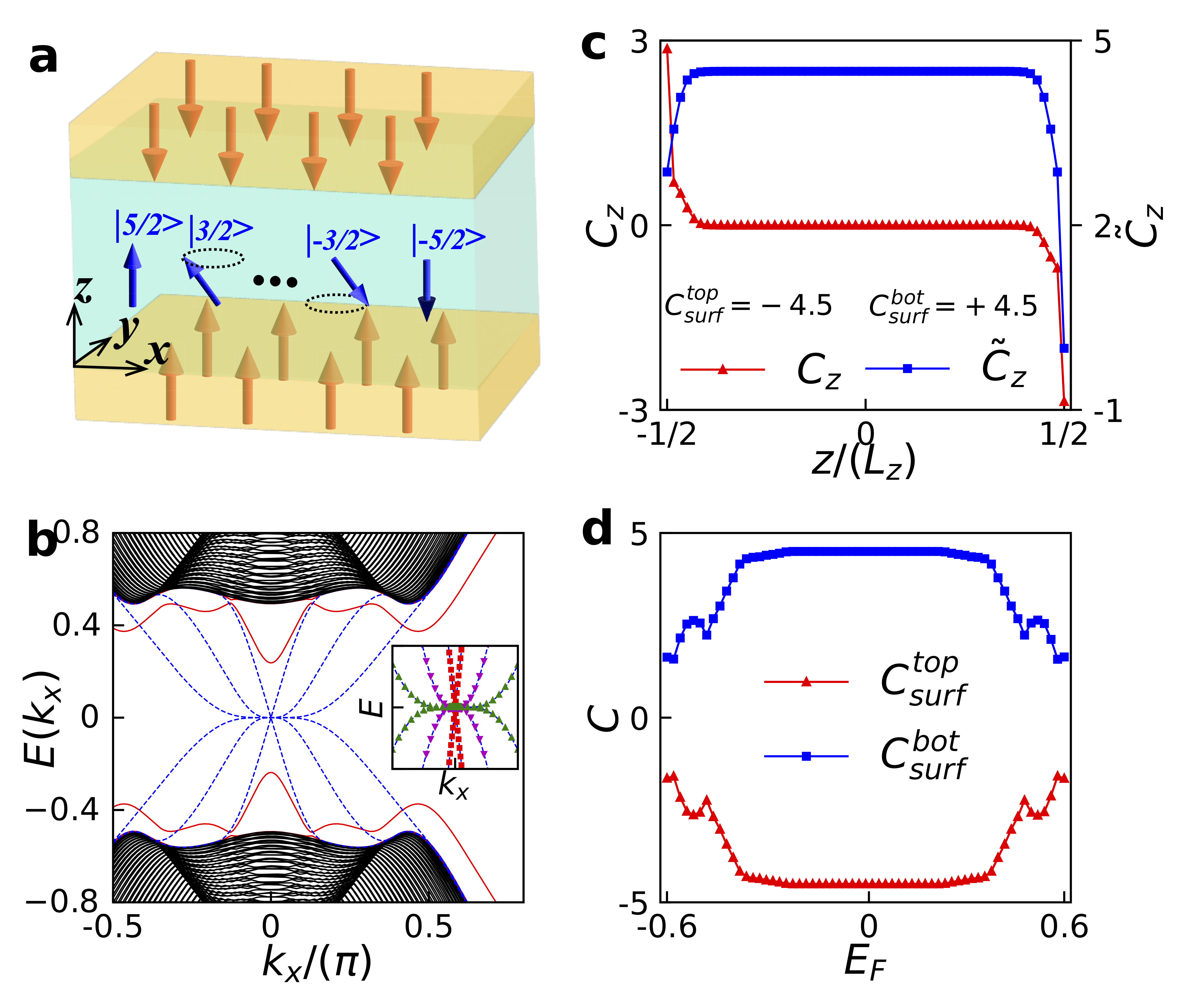}
    \caption{\textbf{Model of the HSAI with spin-$5/2$. a} Schematic for the HSAI with spin-$5/2$. The magnetic quantum number $m_z$ takes values from $-5/2$ to $5/2$.
    	\textbf{b} Two dimensional energy spectra for spin-$5/2$ HSAI  along $k_x$ ($k_y=0$) in the absence (blue dashed lines) and presence (solid magenta lines) of the exchange interaction. The black lines are bulk bands. Inset: Solid lines are energy spectra of the HSAI without magnetic exchange term near the charge neutral point while markers are the fitting data.
	\textbf{c} Layer-resolved Chern number $C(z)$ and the cumulative Chern number $\tilde{C}(z)=\sum_{-1/2}^zC(z)$ versus the layer index $z$.
	\textbf{d} Surface Chern number as a function of the Fermi energy $E_F$. 
	Here, the thickness of the spin-$5/2$ HSAI slab is $L_z=60$ and the Fermi energy is $E_F=0$.
	}
    \label{sfig1}
\end{figure*}
    
Supplementary figure~\ref{sfig1}b shows the energy spectra for the spin-$5/2$ HSAI with and without the exchange term, respectively. In the absence of the exchange term, the system is a spin-$5/2$ topological insulator preserving both time-reversal and space inversion symmetries individually. It has a gaped bulk band but supports three gapless surface bands at each surface. In this case, the three surface bands can be perfectly fitted by a linear Dirac band $E_1\sim k$, a cubic band $E_2\sim k^3$ and a quintuple band $E_3\sim k^5$ as shown in the inset in Supplementary Fig.~\ref{sfig1}b. The exchange term between the magnetic moments and the spin-$5/2$ topological electrons explicitly breaks the time-reversal symmetry of the HSAI and hence opens an exchange surface gap. When the Fermi energy lies inside the band gap, the lay-resolved Chern number $C(z)$ can be derived by projecting the TKNN formula into the specific layer. The results are shown in Supplementary Fig.~\ref{sfig1}c along with the cumulative Chern numbers $\tilde{C}(z)=\sum_{-L_z/2}^zC(z)$. We discover that the layer-resolved Chern numbers lie oppositely inside few surface layers, leading to a vanishing total Chern number $C=0$. However, the surface Chern number in one side is $C^{surf}_{bot(top)}=\pm9/2$, which indicates a distinctive axion term $\theta_{5/2}=(C^{surf}_{bot}-C^{surf}_{top})\pi=9\pi$~\cite{Andrew2009}. This axion term is well preserved as long as the Fermi energy remains inside the band gap as shown in Supplementary Fig.~\ref{sfig1}d.

\begin{figure*}[ht]
    \centering
    \includegraphics[width=0.6\linewidth]{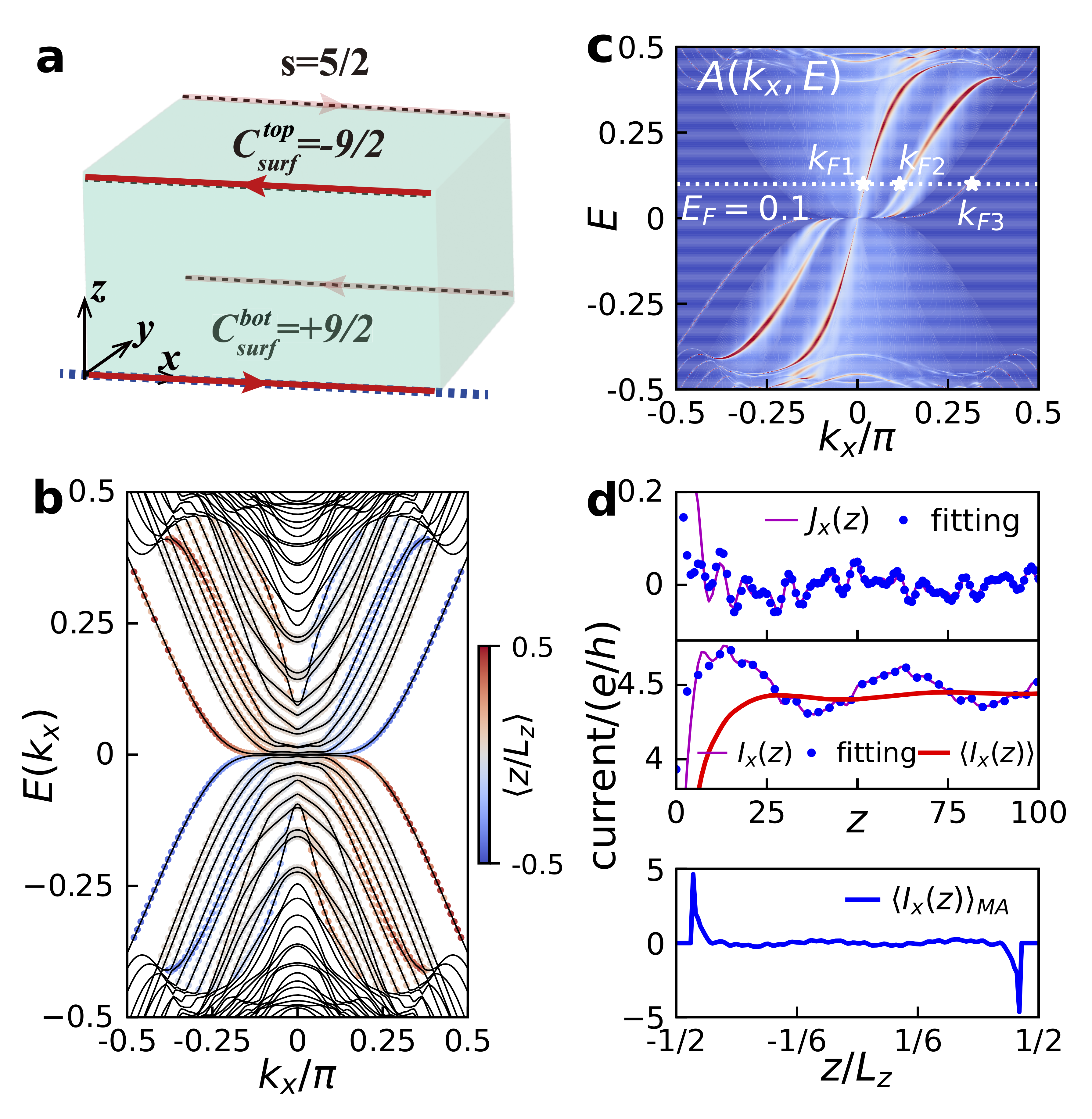}
    \caption{\textbf{Quantized helical hinge current in HSAI with spin-$5/2$. a} Schematic current flow in a spin-$5/2$ HSAI slab. 
    	\textbf{b} One dimensional energy spectrum and the average position $\langle z/L_z\rangle$ on the front surface as functions of $k_x$ with $L_y=20$, $L_z=24$.
	\textbf{c} Spectrum density $A(k_x, E)$ for the front lower hinge as marked by the purple star in \textbf{a} on the $k_x-E$ plane.  Here, the system size is $L_y=20$, $L_z=\infty/2$.
	\textbf{d} Top and middle panels are the current density $J_x(z)$, current flux $I_x(z)$ and its $z$-averaged flux $\langle I_x(z)\rangle$ versus the layer index $z$ for a semi-infinite system with size $L_y=20$, $L_z=\infty/2$. The blue dots are the fitting data using three power law decaying edge currents with momenta $k_{F1}$, $k_{F2}$ and $k_{F3}$ as marked by the white dots in \textbf{c}. Bottom panel shows the distribution of the moving averaged current $\langle I_x(z)\rangle_{MA}$ on the front surface with system size $L_y=20$, $L_z=150$
	}
    \label{sfig2}
\end{figure*}

Even though the surface Chern number in spin-$5/2$ HSAI is not an integer, it still supports an identically quantized helical hinge current due to the difference of the quantum anomalous Hall conductances between top (or bottom) surface and the neighboring side surfaces. To explore this quantized helical hinge current, we examine the average position $\langle z/L_z\rangle$ (Supplementary Fig.~\ref{sfig2}b) as well as the energy spectrum $A(k_x,E)$ (Supplementary Fig.~\ref{sfig2}c) on a HSAI slab. The two results shown in Figs.~\ref{sfig2}b and c confirm a pair of helical hinge currents on the front surface of the HSAI slab at $y=-L_y/2$. Because of the inherent $\mathcal{PT}$ symmetry, two additional helical hinge currents exist on the other surface at $y=L_y/2$. The diagonal hinge currents are connected by this $\mathcal{PT}$ symmetry. They much propagate along the same direction. Therefore, in analogy to that in spin-$3/2$ HSAI and spin-$1/2$ axion insulator, the spin-$5/2$ HSAI also support helical hinge currents that propagate oppositely on the neighboring hinges. To further quantitatively uncover those helical hinge currents, we calculate the current density $J_x(z)$, current flux $I_x(z)$, $z$-average current $\langle I_x(z)\rangle$ and the moving averaging current flux $\langle I_x(z)\rangle_{MA}$ in accordance with the formalism provided in the Methods. The results are plotted in Supplementary Fig.~\ref{sfig2}d. On one hand, $J_x(z)$ and $I_x(z)$ (solid magenta lines in the top and middle panels) coincide remarkably well with the fitting data (blue dots) obtained from the superposition of three power law decaying edge currents $J_x^{i=1,2,3}=a_i\cos{(2k_{Fi}a_0+\phi_i)}/\sqrt{z}$, where $k_{Fi}$ refers to the Fermi momenta marked by the white dots in Supplementary Fig.~\ref{sfig2}c, $a_i$ and $\phi_0$ are fitting parameters, certifing that the beating mode of $J_x(z)$ comes from the superposition of the coherent edge currents on the same hinge. This beating mode occurs only in the HSAI since it originates from the superposition of two or more coherent edge currents while spin-$1/2$ axion insulator harbors only one edge current. On the other hand, one can see that $\langle I_x(z)\rangle$ and $\langle I_x(z)\rangle_{MA}$ quantize to $\pm4.5e/h$ only a few layer away from the hinge. Thus, those signatures evidently confirms the quantized helical hinge current identical to the surface Chern number.

 \begin{figure*}[ht]
    \centering
    \includegraphics[width=0.6\linewidth]{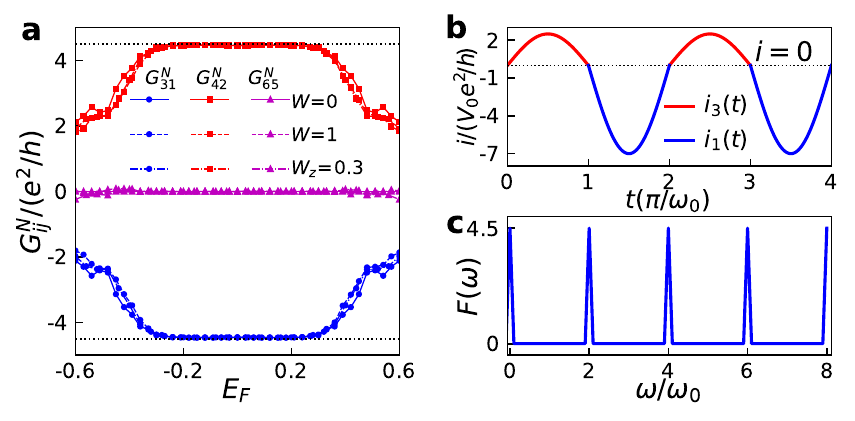}
    \caption{\textbf{Transport properties of the spin-$5/2$ HSAI in a six terminal device. a} Ensemble-averaged non-reciprocal conductances versus the Fermi energy in the clean limit, with non-magnetic Anderson disorders $W=1$ and with magnetic Anderson disorders $W_z=0.3$. Here, the system size is $L_x = 31$, $L_y = 20$, $L_z = 21$, and the size of transverse terminals is $10\times 10$.
    	\textbf{b} Corresponding temporal dependent current output between terminals 1 and 3 with parameters $G_{13}=2.5e^2/h$ and $G_{31}=7e^2/h$.
	\textbf{c} $F(\omega)$ as a function of the frequency $\omega$.
	}
    \label{sfig3}
\end{figure*}

In the same six terminal device as schematically shown in Fig. 2e in the main text, we calculate the non-reciprocal conductance for the spin-$5/2$ HSAI by using the non-equilibrium Green's function method. Three representative non-reciprocal conductances $G_{31}^N$, $G_{42}^N$ and $G_{65}^N$ as functions of the Fermi energy $E_F$ are plotted in Supplementary Fig.~\ref{sfig3}a under different conditions (in the clean limit $W=0$, with non-magnetic Anderson disorders $W=1.0$ and with magnetic Anderson disorders $W_z=0.3$). We see that $G_{31}^N=4.5e^2/h$, $G_{42}^N=-4.5e^2/h$ and $G_{65}^N=0$ when the Fermi energy lie inside the band gap regardless of the presence of disorders, confirming that those quantized non-reciprocal conductances are chiral and robust against both non-magnetic and magnetic disorders. As a result, the quantized helical hinge current is topological protected. This quantized helical hinge current can be measured by using the alternate detection method, which eliminates the contribution from the conductive side surface. During the first half period, a positive voltage is applied to terminal 1, while current flows from terminal 3 is detected [$i_3(t)$]. Owing to the opposite chirality, the hinge current in this case is completely blocked, allowing only the current on the side surface to flow from terminal 1 to terminal 3. In the second half period, the harmonic voltage is moved to terminal 3, and the current output $i_1(t)$ from terminal 1 includes contributions from both the side surface and the hinge. Consequently, the temporal dependent current $i(t)=i_1(t)+i_3(t)$ is asymmetric as shown in Supplementary Fig.~\ref{sfig3}b. The amplitude difference between the positive and negative current originates from the chirality of quantized helical hinge current. This difference can be revealed quantitatively by the function $F(\omega)$ as explained in Sec.~\ref{alter_detect}, which equals to the non-reciprocal conductance $G^N_{13}=G_{13}-G_{31}$ when $\omega=2\omega_0$ (Supplementary Fig.~\ref{sfig3}c). This further affirms that the quantized helical hinge current in HSAI with different spins can be experimentally observed via the alternate detection proposed in the main text.

 \begin{figure*}[ht]
    \centering
    \includegraphics[width=0.98\linewidth]{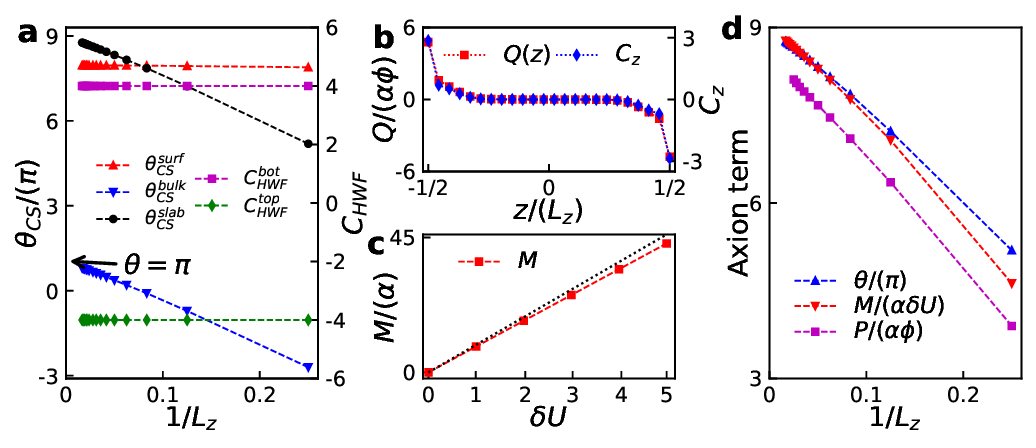}
    \caption{\textbf{Axion term and the topological magneto-electric effect in HSAI with spin-$5/2$. a} Axion terms and the surface Chern numbers versus the inverse layer thickness obtained by using the HWFs.
    	\textbf{b}  Magnetic field induced charge distribution along $\hat{z}$-direction and the layer-resolved Chern number for a spin-$5/2$ HSAI. Here, the charge polarization is obtained on a HSAI slab with open boundary condition along $\hat{y}$-direction ($L_y=40$) but periodic boundary condition along $\hat{x}$-direction.
	\textbf{c} Electric field induced orbital magnetization for a spin-$5/2$ HSAI. The black dashed line shows the ideal case (IC) with an exact axion term $\theta=9\pi$.
	\textbf{d} Size scaling of the axion term $\theta_{CS}^{slab}/\pi$, polarization coefficient $P/(\alpha\phi)$, and magnetization coefficient $M/(\alpha\delta U)$.
	Here, the system size is $L_z=24$.
	}
    \label{sfig4}
\end{figure*}

We next examine the axion term and the topological magneto-electric effect bonded to it. The axion term can be calculated in terms of the hybrid Wannier functions on a two dimension HSAI slab with thickness $L_z$. Supplementary figure~\ref{sfig4}a shows the surface axion term $\theta_{CS}^{surf}$, bulk axion term $\theta_{CS}^{bulk}$, and the total axion term $\theta_{CS}^{slab}$ as functions of the inverse layer thickness $1/L_z$. It shows that the total axion term $\theta_{CS}^{slab}$ of the spin-$5/2$ HSAI is $9\pi$ when the system size approaches infinity while the bulk axion term quantizes to $\theta_{CS}^{bulk}=\pi$ simultaneously. Therefore, in analogy to the spin-$1/2$ case, the spin-$5/2$ HSAI is an axion insulator protected by the $\mathcal{PT}$ symmetry in the bulk. The surface axion term corresponding from the surface Wannier function is $\theta_{CS}^{surf}=8\pi$, which is consistent with the Chern number of the surface Wannier function shown in the same figure and is independent of the system size. As the surface Wannier functions exist only on a slab geometry, they are independent of the system size and disappear under the periodic boundary condition. Thus, the axion term in the bulk for the spin-$5/2$ HSAI is $\pi$, which equals to $\theta_{CS}^{bulk}$ obtained here on a slab geometry when the system size is infinity. This quantized axion term is also protected by the $\mathcal{PT}$ symmetry in analogy to the spin-$1/2$ axion insulator.
The non-vanishing axion term of the slab indicates an identical topological magneto-electric effect.
In the presence of an external magnetic field, the electrons can be pushed from one side to the other as plotted in Supplementary Fig.~\ref{sfig4}b, resulting in a charge polarization with the coefficient $P/(\alpha\phi)$ identical to the axion term $\theta_{CS}^{slab}$, where $P$ is the charge polarization, $\alpha$ is the fine structure constant and $\phi$ the total magnetic flux penetrating the HSAI slab. Furthermore, the charge distribution is consistent with the layer-resolved Chern numbers $C(z)$ obtained from the Bloch wave functions as shown in the same figure. By contrast, the electric field induced magnetization (red squares) shown in Supplementary Fig.~\ref{sfig4}c also agree quantitatively well with the ideal case with an exact quantized axion field $\theta_{IC}=9\pi$ (black line), in which the slope of the data $M/(\alpha \delta U)$ refers to the magnetization coefficient with $\delta U$ the total potential drop across the HSAI slab. The slight deviation between them can be ascribed to the finite size effect, which is further unveiled by the size scalings of the axion term $\theta_{CS}^{slab}/\pi$, polarization coefficient $P/(\alpha\phi)$, and magnetization coefficient $M/(\alpha\delta U)$ shown in Supplementary Fig.~\ref{sfig4}d. The results establish the equivalence of the slab axion term, the polarization coefficient and the magnetization coefficient.

 \begin{figure*}[ht]
    \centering
    \includegraphics[width=0.6\linewidth]{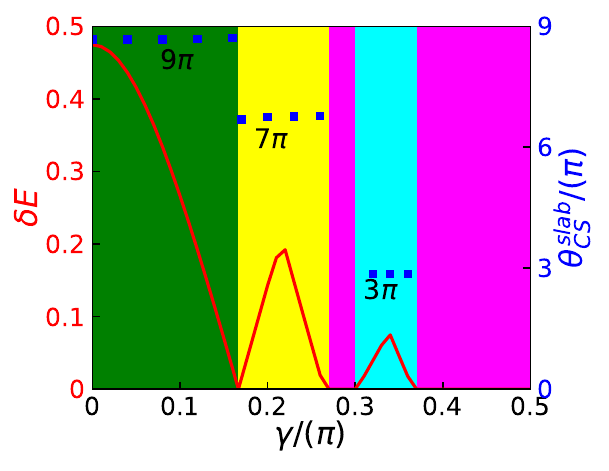}
    \caption{\textbf{Topological phase transition in HSAI with spin-$5/2$.} The red solid line plots the band gap for the HSAI with spin-$5/2$ as a function of the canting angle $\gamma$ while the blue square refers to the corresponding axion term $\theta_{CS}^{slab}$. The axion term is $\theta_{CS}^{slab}=9\pi$ in the green region, $\theta_{CS}^{slab}=7\pi$ in the yellow region, and $\theta_{CS}^{slab}=3\pi$ in the cyan region. In the magenta regions, the gap is closed, therefore the axion term is not well-defined since the system is a metal. Here, the system size is $L_z=40$. 
	}
    \label{sfig5}
\end{figure*}

Finally, we explore the topological phase transition in spin-$5/2$ HSAI under the driven of an in-plane magnetic field, which turns the antiparallel spins into canted spin state. The red line in Supplementary Fig.~\ref{sfig5} plots the two dimensional energy band gap for the spin-$5/2$ HSAI as a function of the canting angle $\gamma$. It exhibits more gap closing and reopen than the spin-$3/2$ HSAI in the main text, which indicates much rich topological phase transitions. The blue squares in the same figure show the corresponding axion terms obtained using the hybrid Wannier function. We observe that the axion term changes from $9\pi$ (green region) to $7\pi$ (yellow region) then to $3\pi$ (cyan region) when the canting angle $\gamma$ is enlarged. Moreover, there exist two regions (magenta regions) where the band gap vanishes. In theses regions the system is a metal, therefore the axion term is not well-defined~\cite{Qi2008}.
It is also important to note that the topological phase transition originates solely from the surface axion term $\theta_{CS}^{surf}$ while the bulk axion term $\theta_{CS}^{bulk}$ remains unchanged during the process.
Because the transport signals in HSAI such as the non-reciprocal conductance identical to the quantized helical hinge currents and the topological magneto-electric response are proportional to the axion field, the HSAI thus provides a platform to realized the goal of axionic topological phase transition.

{\section{Even-odd effect of the high spin axion insulator}\label{even_odd}}

\begin{figure*}[ht]
    \centering
    \includegraphics[width=0.7\linewidth]{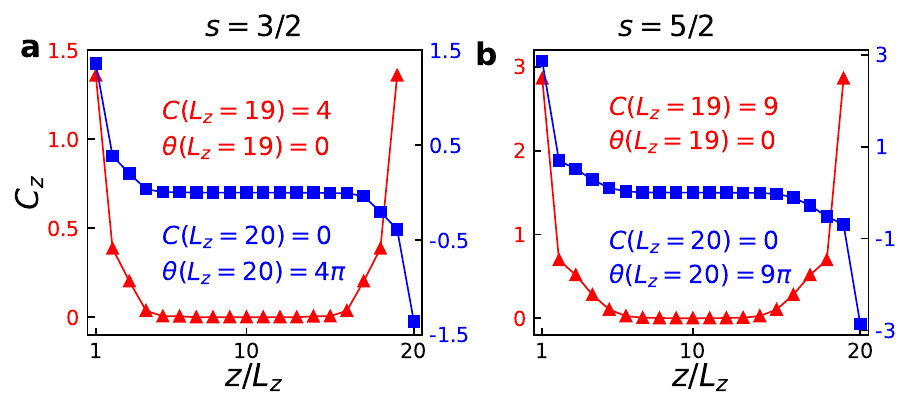}
    \caption{\textbf{Even-odd effect of HSAI.} \textbf{a} and \textbf{b} are layer-resolved Chern number for spin-$3/2$ and spin-$5/2$ HSAI with different layer thicknesses $L_z=19$ (red triangle) and $L_z=20$ (blue square). The Chern numbers and axion fields are labeled in the figure.
	}
    \label{sfig6}
\end{figure*}

In the odd layer system, the net magnetization is non-vanishing because of the uncompensated magnetic layer. We can thus simulate this state by using parallel magnetic moments on both top and bottom layers. In order to reveal the difference between them, we explore the layer resolved Chern numbers, which can be obtained by using Eq. (4) in the methods. The results displayed in Supplementary Fig.~\ref{sfig6} show that the layer-resolved Chern numbers for odd layer systems distribute symmetrically on the top and bottom layers, leading to a vanishing axion field $\theta=0$ while a nonzero Chern number $C=(s+1/2)^2$ analogous to the odd layer MnBi${}_2$Te${}_4$~\cite{Lin2022}. Therefore, the odd layer system is a high Chern number insulator. On the contrary, the even layer system is a HSAI with an axion field $\theta=(s+1/2)^2\pi$ because of the asymmetric layer-resolved Chern numbers.